\documentclass[aps,prd,preprint,floatfix]{revtex4}
\usepackage{epsfig}
\usepackage{bm}

\begin{document}
 
\preprint{\rightline{ANL-HEP-PR-06-33}}
 
\title{Quantization and simulation of Born-Infeld non-linear electrodynamics on
a lattice}
 
\author{J.~B.~Kogut}\thanks{Supported in part by NSF grant NSF PHY03-04252.}
\address{Department of Energy, Division of High Energy Physics, Washington, DC 
20585, USA\\
and\\
Dept. of Physics -- TQHN, Univ. of Maryland, 82 Regents Dr., College
Park, MD 20742, USA}
\author{D.~K.~Sinclair}\thanks{This work was supported by the U.S.
Department of Energy, Division of High Energy Physics, Contract \\*[-0.1in]
W-31-109-ENG-38.}
\address{HEP Division, Argonne National Laboratory, 9700 South Cass Avenue,
Argonne, IL 60439, USA}
 
\begin{abstract}
Born-Infeld non-linear electrodynamics arises naturally as a field theory
description of the dynamics of strings and branes. Most analyses of this
theory have been limited to studying it as a classical field theory. We
quantize this theory on a Euclidean 4-dimensional space-time lattice and
determine its properties using Monte-Carlo simulations. The electromagnetic
field around a static point charge is measured using L\"{u}scher-Weisz methods
to overcome the sign problem associated with the introduction of this charge.
The $\bf{D}$ field appears identical to that of Maxwell QED. However, the
$\bf{E}$ field is enhanced by quantum fluctuations, while still showing the
short distance screening observed in the classical theory. In addition,
whereas for the classical theory, the screening increases without bound as the
non-linearity increases, the quantum theory approaches a limiting conformal
field theory.
\end{abstract}
\maketitle

\section{Introduction}

Theories of strings and branes have shown much promise as theories unifying
gravity with strong, weak and electromagnetic interactions. One limitation
has been that most of the work on these theories has been performed on the
classical theories. The most promising way to produce quantum theories of
strings and branes has been to study field theories which describe their
dynamics. The field theories which arise most naturally in this context are
Born-Infeld non-linear electrodynamics \cite{Born:1934,Born:1934gh}
and its extensions. Early papers which use Born-Infeld theories to describe the
dynamics of strings and branes include \cite{Fradkin:1985qd,Leigh:1989jq,
Aganagic:1996nn,Gibbons:1997xz,Tseytlin:1999dj}. However, since much of the
interesting physics arising from the non-linearities in Born-Infeld models is
inaccessible to perturbation theory, most of the work on these theories has
also been limited to the classical domain \cite{Callan:1997kz,Gibbons:1997xz}
or to the small quantum fluctuations \cite{Rey:1998ik} around these classical
solutions. Beyond this, there have only been a few exploratory investigations
of the general problem of quantizing this theory, for example
\cite{Hotta:2004zj}, which examines quantum solutions representing propagation
in a fixed direction. 

Born-Infeld electrodynamics in $n+1$ dimensions is described by the non-linear
action
\begin{equation}
S = b^2 \int d^{n+1}x\,\left[1-\sqrt{-\det\left(g_{\mu\nu}+
\frac{1}{b}F_{\mu\nu}\right)}\,\right].
\label{eqn:Sminkowski}
\end{equation}
To make the connection with strings and branes, one typically chooses $n=9$
and dimensionally reduces the theory from $9+1$ down to $p+1$ dimensions. This
describes a $p$-brane, and the extra $9-p$ components of $A_\mu$ are
identified with the transverse degrees of freedom of the brane. The special
case $p=1$ describes a string. Here the non-linearity parameter $b$ is related
to the string tension through $1/b=2\pi\alpha'$.

In this work we study the original Born-Infeld theory in $3+1$ dimensions
($n=3$) with no dimensional reduction. In brane language we study a $3$-brane
with no transverse dimensions. This theory was originally proposed as a
modification of QED in which the electric field of a static point particle
was screened at short distances, rendering the self-energy of a point particle
finite \cite{Born:1934,Born:1934gh}.

We note that, in Euclidean space, the action is positive so that quantization
by the standard functional integral techniques is well defined. We then define
this theory on a discrete space-time lattice, preserving gauge invariance. The
positivity of the action allows the use of Metropolis Monte-Carlo 
\cite{Metropolis:1953am} simulations to extract the properties of the quantum
field theory. In particular, we measure the electromagnetic fields produced by
a static point charge by including the effects of a Wilson Line (Polyakov
Loop) in the action. This destroys the positivity, but we are able to use the
methods of L\"{u}scher and Weisz \cite{Luscher:2001up} (after Parisi,
Petronzio and Rapuano \cite{Parisi:1983hm}) to produce the exponential
statistics required to overcome this lack of positivity.

We have performed simulations of the lattice Born-Infeld theory on $8^4$ and 
$12^4$ lattices. We present evidence that the ${\bf D}$ field emanating from a 
point charge is identical with that of the Maxwell theory, for all values of
$b$, up to lattice artifacts. The ${\bf E}$ field shows short-distance screening
as for the classical theory, but whereas classically the screening length
increases without limit as $b \rightarrow 0$, the quantum field theory
approaches a conformal field theory in this limit. In addition, the ${\bf E}$
field is enhanced over its classical value by quantum fluctuations. Preliminary
results for the smaller lattice were presented at Lattice 2005, Dublin
\cite{Sinclair:2005yw}.

In section 2 we summarise the salient features of the classical theory from
the extensive literature on Born-Infeld electrodynamics. Section 3 discusses
the lattice formulation in Euclidean space and simulation methods. We present
the results of our simulations in section 4. These results are discussed and
conclusions drawn in section 5.

\section{Classical Born-Infeld electrodynamics}

In this section we summarise those results from the literature on the classical
Born-Infeld theory in $3+1$ dimensional Minkowski space which are relevant to
our investigations. These results are condensed from the following references
\cite{Bialynicki-Birula:1984tx,Chruscinski:1997pe,Chruscinski:1997jd,%
Chruscinski:1998uw}. Evaluating the determinant in equation
\ref{eqn:Sminkowski} leads to the Lagrangian
\begin{eqnarray}
{\cal L} &=& b^2 \left[1-\sqrt{1+\frac{1}{2b^2}F_{\mu\nu}F^{\mu\nu}-
\frac{1}{16b^4}(F_{\mu\nu}\tilde{F}^{\mu\nu})^2}\right]     \nonumber      \\
&=& b^2 \left[1-\sqrt{1-b^{-2}(\bm{E}^2-\bm{B}^2)-b^{-4}(\bm{E \cdot B})^2}
\right].
\end{eqnarray}
Note that the requirement that the argument of the square root be positive,
restricts the magnitude of ${\bf E}$ which in turn gives rise to short-distance
screening and leads to a finite self energy for a point charge. This was the
original reason Born and Infeld proposed this non-linear extension of 
electrodynamics. When dimensionally reduced Born-Infeld theories are used to
describe strings or branes, this same mechanism restricts the energy density of
the string/brane to be finite.
 
To proceed, it is customary to define ${\bf D}$ and ${\bf H}$ fields by
\begin{eqnarray}
\bm{D} &=& {\partial{\cal L} \over \partial\bm{E}} =
               {\bm{E}+b^{-2}(\bm{E\cdot B})\bm{B} \over
               \sqrt{1-b^{-2}(\bm{E}^2-\bm{B}^2)-b^{-4}(\bm{E\cdot B})^2}}
\nonumber \\
\bm{H} &=& {\partial{\cal L} \over \partial\bm{B}} =
               {\bm{B}-b^{-2}(\bm{E\cdot B})\bm{E} \over
               \sqrt{1-b^{-2}(\bm{E}^2-\bm{B}^2)-b^{-4}(\bm{E\cdot B})^2}}.
\label{eqn:DH}
\end{eqnarray}
In terms of ${\bf E}$, ${\bf D}$, ${\bf B}$ and ${\bf H}$, the field equations
are the standard Maxwell equations
\begin{eqnarray}
\nabla \bm{\cdot D} & = & \rho     \nonumber                                \\
\nabla \bm{\cdot B} & = & 0        \nonumber                                \\
\nabla \times \bm{E} + {\partial \bm{B} \over \partial t} & = & \bm{0} 
\nonumber \\
\nabla \times \bm{H} - {\partial \bm{D} \over \partial t} & = & \bm{j},
\end{eqnarray}
where, as usual, the interaction with the electromagnetic current $j_\mu$ is
introduced by adding a term $j^\mu A_\mu$ to the Lagrangian. All the 
non-linearity is hidden in the defining relations of equations~\ref{eqn:DH}. 

For a static point charge $\rho=e\delta^3(\bm{r})$, 
$\nabla \bm{\cdot D} = \rho$ has the spherically symmetric solution
\begin{equation}
\bm{D}={e \over 4\pi r^3}\bm{r}
\end{equation}
as for the Maxwell case. $\bm{B}=\bm{H}=\bm{0}$ and equations~\ref{eqn:DH} then
yield
\begin{equation}
\bm{E}={e \over 4\pi r}{\bm{r} \over \sqrt{r^4+r_0^4}}
\end{equation}
where
\begin{equation}
r_0=\sqrt{|e| \over 4\pi b}
\end{equation}
defines the screening length. Due to this short distance screening, the electric
field at the origin has magnitude $b$ rather than being infinite, and the energy
in the field of the point charge is finite.
 
\section{Euclidean Born-Infeld electrodynamics on a lattice}

In Euclidean space, the Born-Infeld action is
\begin{eqnarray}
S &=& b^2\int d^4x \left[\sqrt{1+\frac{1}{2b^2}F_{\mu\nu}F_{\mu\nu}+
     \frac{1}{16b^4}(F_{\mu\nu}\tilde{F}_{\mu\nu})^2}-1\right] \nonumber     \\
  &=& b^2\int d^4x \left[\sqrt{1+b^{-2}(\bm{E}^2+\bm{B}^2)
                         +b^{-4}(\bm{E \cdot B})^2}-1\right].
\end{eqnarray}
We note that this action is positive, so this theory can be simulated using
importance sampling methods. The expressions for ${\bf D}$ and ${\bf H}$ are
identical to those of equations~\ref{eqn:DH}, except that all signs are
positive.

To enable simulations, we must first transcribe this field theory to a discrete
space-time lattice. We choose a hypercubic lattice with lattice spacing $a$.
For convenience we define $\beta=b^2a^4$. From now on we will work in lattice
units where $a$ is set to $1$, except where we discuss the effects of varying
$a$. Because power counting suggests this theory is non-renormalizable, it is
generally considered as an effective theory, requiring a cutoff. Hence we are
not required to take the $a \rightarrow 0$ limit. We shall have more to say 
about this later.

We choose the non-compact formulation of QED on the lattice, since this is
closest to continuum QED (In the Maxwell case it defines a solvable, free field 
theory, as in the continuum). This is done by defining the lattice $F_{\mu\nu}$
by 
\begin{equation}
F_{\mu\nu}(x+{\textstyle \frac{1}{2}}\hat{\mu}
            +{\textstyle \frac{1}{2}}\hat{\nu})
= A_\nu(x+\hat{\mu})-A_\nu(x)-A_\mu(x+\hat{\nu})+A_\mu(x),
\end{equation}
which is gauge invariant as is the continuum $F_{\mu\nu}$. This indicates the
first subtlety; $F_{\mu\nu}$ is not associated with a single site $x$ or link
of the original lattice. Therefore to preserve the symmetries of the cubic
lattice we average our action over the 16 choices of $6$ plaquettes associated
with a lattice site. These 16 choices are defined by choosing 1 unit vector in
each of the 4 directions emanating from the site, in all possible ways. The
set of plaquettes associated with a given choice of unit vectors is uniquely
defined by requiring that 2 edges of each plaquette belong to the chosen set
of unit vectors. Simulations are performed using Metropolis Monte-Carlo
updates of the gauge fields ($A_\mu$) on each of the links of the lattice. We
periodically perform subtractions of constant fields from each $A_\mu$, since
such constant fields have no physics, to prevent the lattice average of $A_\mu$
becoming too large. Similarly, we periodically gauge-fix these fields to Landau
gauge, to remove large gauge fluctuations.

To compare with the classical Born-Infeld results we need to measure the
${\bf E}$ and ${\bf D}$ fields for a point charge. Since we use periodic
boundary conditions on the gauge fields, the total charge on the lattice must
be zero. We choose the `Jellium' solution from condensed matter physics to
circumvent this difficulty. Here, if the magnitude of the point charge is $e$,
we distribute a charge $-e$ uniformly over the lattice. Hence each site
has charge $-e/{\cal V}$ (${\cal V}=N_xN_yN_z$ is the spatial volume of the
lattice), except for that site containing the point charge, whose charge is now
reduced to $e-e/{\cal V}$. On the lattice this charge is introduced by 
including a Wilson Line (Polyakov loop) $W(\bm{x})$ in the functional integral.
\begin{equation}
W(\bm{x}) = \exp\left\{i e \sum_t \left[A_4(\bm{x},t)
          - \frac{1}{\cal V}\sum_{\bm{y}} A_4(\bm{y},t)\right]\right\}.
\end{equation}
We calculate the $\langle\bm{E}\rangle$ and $\langle\bm{D}\rangle$ due to this
charge as
\begin{eqnarray}
i \langle\bm{E}\rangle_\rho (\bm{y}-\bm{x}) &=&
{\langle\bm{E}(\bm{y},t)W(\bm{x})\rangle \over \langle W(\bm{x})\rangle}
\nonumber \\
i \langle\bm{D}\rangle_\rho (\bm{y}-\bm{x}) &=&
{\langle\bm{D}(\bm{y},t)W(\bm{x})\rangle \over \langle W(\bm{x})\rangle}.
\label{eqn:expect}
\end{eqnarray}
where the $i$ converts our electric fields from Euclidean space to Minkowski
space, and we use the subscript $\rho$ to denote the expectation value in the
presence of charge density $\rho$.

The Wilson Line for fixed ${\bf x}$ on any given configuration has magnitude 
one, so
\begin{equation}
\langle |W(\bm{x})|\rangle = 1,
\end{equation}
while it is expected that
\begin{equation}                                                            
\langle W(\bm{x}) \rangle = e^{-const\: N_t}.                      
\end{equation} 
This indicates that the phase of $W$ causes important cancellations between
the Wilson lines for different configurations in our ensemble. Hence we have
a sign problem. Averaging over the site ${\bf x}$ helps improve our statistics.
Even if we measured the Wilson line averaged over all sites, every sweep, we
will run into trouble if
\begin{equation}                                                            
\langle W(\bm{x})\rangle\lesssim 1/\sqrt{(number\: of\: sweeps \times {\cal V}},
\label{eqn:statistics}
\end{equation}   
in the most optimistic scenario where the phases of the Wilson lines on a given
configuration are uncorrelated as are those on consecutive sweeps.
For any of our $12^4$ runs, the number of sweeps is 500,000 and ${\cal V}=1728$
and the right hand side of equation~\ref{eqn:statistics} is 
$\approx 3.4 \times 10^{-5}$. As we shall see later, we really need to do 
better.

L\"{u}scher and Weisz \cite{Luscher:2001up} (following Parisi, Petronzio and 
Rapuano \cite{Parisi:1983hm} ) have pointed out that one can effectively
enhance ones statistics by dividing the lattice into some number (call it $n$)
of timeslices. Notice that, if one fixes the link fields on the boundaries of
these timeslices, the Monte Carlo updates of those fields on the interiors of
each slice are independent of the updates of the interiors of all other
slices. Thus if one performs $m$ updates of the interior of each time-slice,
it is equivalent to performing $m^n$ (restricted) updates of the lattice. In
addition this process can be performed recursively over progressively coarser
slicings of the lattice. In practice we have used thickness $1$ and $2$ time
slices recursively.

Let us now examine the limiting cases, first where $b \rightarrow \infty$, and
second where $b \rightarrow 0$. As $b \rightarrow \infty$, the Euclidean
Lagrangian approaches its Maxwell (free field) form
\begin{equation}
{\cal L}_E = \frac{1}{2}(\bm{E}^2 + \bm{B}^2).
\end{equation}
In the limit $b \rightarrow 0$ the Euclidean Lagrangian approaches the limiting
form
\begin{equation}
{\cal L}_E = |\bm{E\cdot B}|,
\label{eqn:limit}
\end{equation}
which describes a 4-dimensional conformal field theory. [Note that the
Minkowski Lagrangian (equation~\ref{eqn:Sminkowski}) does not exist in this
limit. However, the Hamiltonian formulation is valid in this limit, and leads
to the Hamiltonian
\begin{equation}
{\cal H} = |\bm{D}\times \bm{B}|,
\end{equation}
which also describes a conformal field theory 
\cite{Bialynicki-Birula:1984tx,Chruscinski:2000zm}.]

\section{Simulation details and results}

We have simulated lattice Born-Infeld QED on $8^4$ and $12^4$ lattices. On
both $8^4$ and $12^4$ lattices we performed runs of 500,000 10-hit Metropolis
sweeps of the lattice at each of $\beta=0.0001,0.01,1,100$. On the $8^4$ 
lattice we ran an additional 100,000 sweeps at each of $\beta=0.1,0.2,0.5,2,5$.
Measurements were made every 100 sweeps. On the $8^4$ lattices the Wilson lines
and the on-axis fields they produced were measured in the L\"{u}scher-Weisz
scheme where we used 10 10-hit Metropolis updates of the interiors of each 
thickness-1 timeslice for each of 10 10-hit Metropolis updates of the
interiors of each thickness-2 timeslice. For the $12^4$ lattice we increased
the number of updates at each level to 20. This means that we effectively used
$10^{12}$ configurations every 100 sweeps for our $8^4$ lattice simulations and
$2.62144 \times 10^{23}$ configurations every 100 sweeps for our $12^4$ runs,
for our measurements of the Wilson lines and electric fields. Admittedly these
configurations generated in the L\"{u}scher-Weisz scheme are not all
independent.

\begin{figure}[htb]
\epsfxsize=6in
\centerline{\epsffile{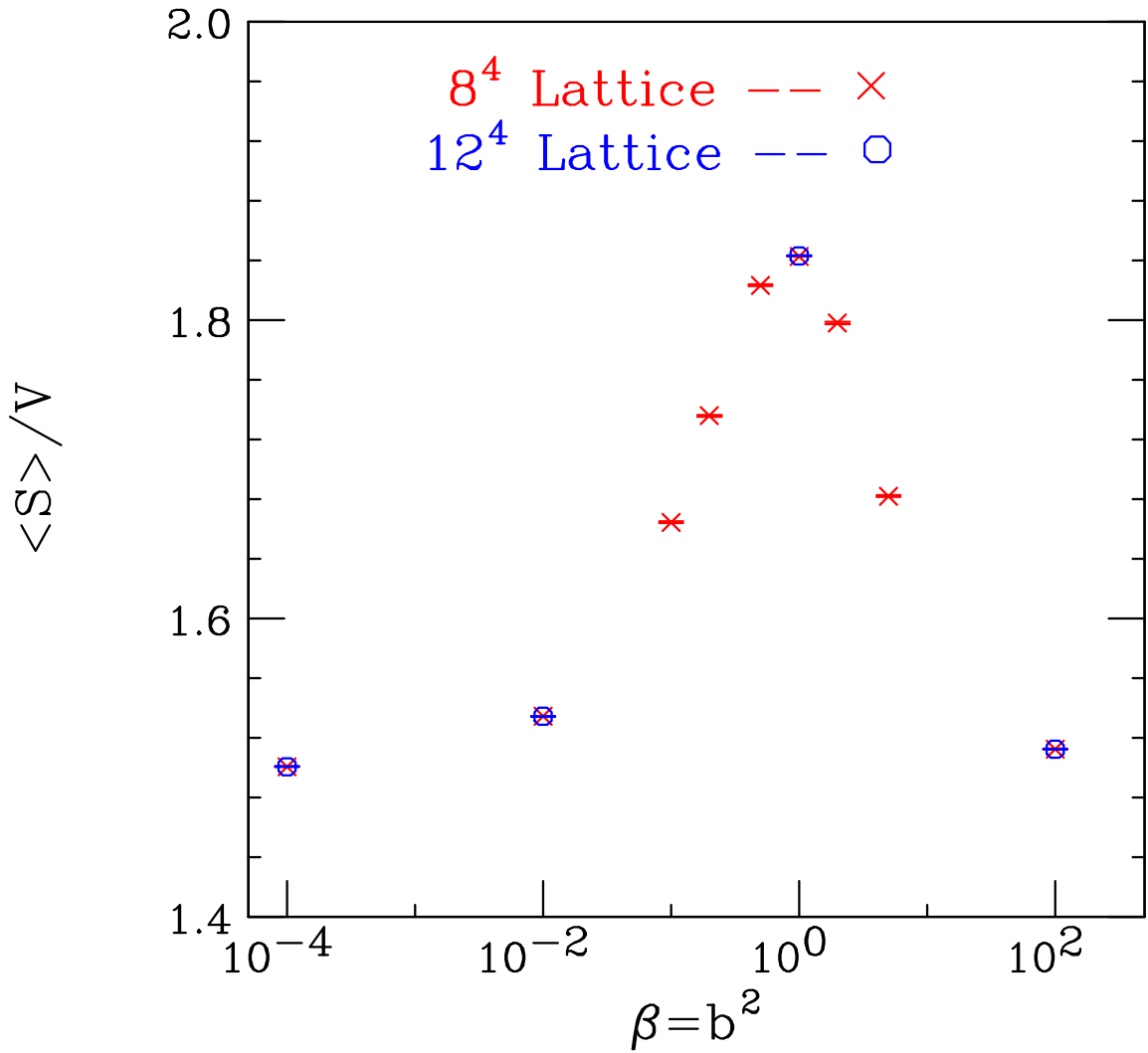}}
\caption{Euclidean action density $\langle S \rangle/V$ as a function of
$\beta$}
\label{fig:energy}
\end{figure}

In figure~\ref{fig:energy} we present the average action density 
$\langle S \rangle/V$, where $V$ is the space-time volume, as a function of
$\beta=b^2a^4$ for our runs on both $8^4$ and $12^4$ lattices. We notice that
there is good agreement between the 2 lattice sizes, indicating that that the
finite volume effects are small. As $\beta \rightarrow \infty$, where the
theory becomes the linear Maxwell theory, the equipartition theorem predicts
that the action density is $\frac{3}{2}$. We note that the action density is
within a percent of this value by $\beta=100$, indicating that for $\beta
\gtrsim 100$ the theory is close to the Maxwell theory. In the 
$\beta \rightarrow 0$ limit the scaling properties of the action again lead to
an action density of $\frac{3}{2}$. We see that our $\beta=0.0001$ action
density is very close to this value, while the $\beta=0.01$ value is only a
little over $2\%$ higher. This suggests that lattice Born-Infeld QED is close
to the limit described by equation~\ref{eqn:limit} for $\beta \lesssim 0.01$.
[Note that, because the zero modes do not contribute to the action, the
limiting values ($\frac{3}{2}$) have corrections ${\cal O}(1/V)$]. In between
the 2 limits the action rises to a maximum of magnitude $<2$ for $\beta \sim 1$.

\begin{figure}[htb]
\epsfxsize=4in
\centerline{\epsffile{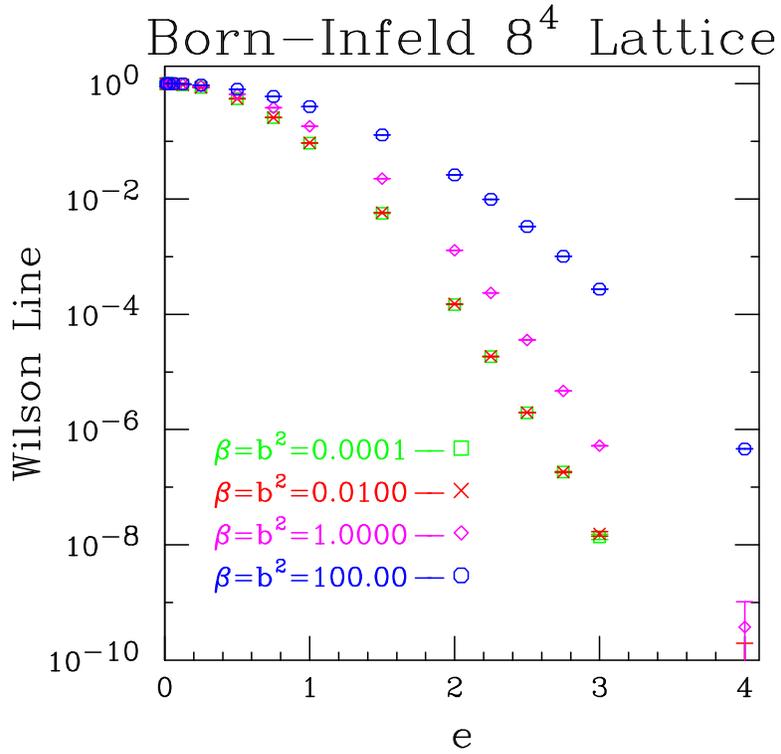}}
\vspace{0.25in}
\centerline{\epsffile{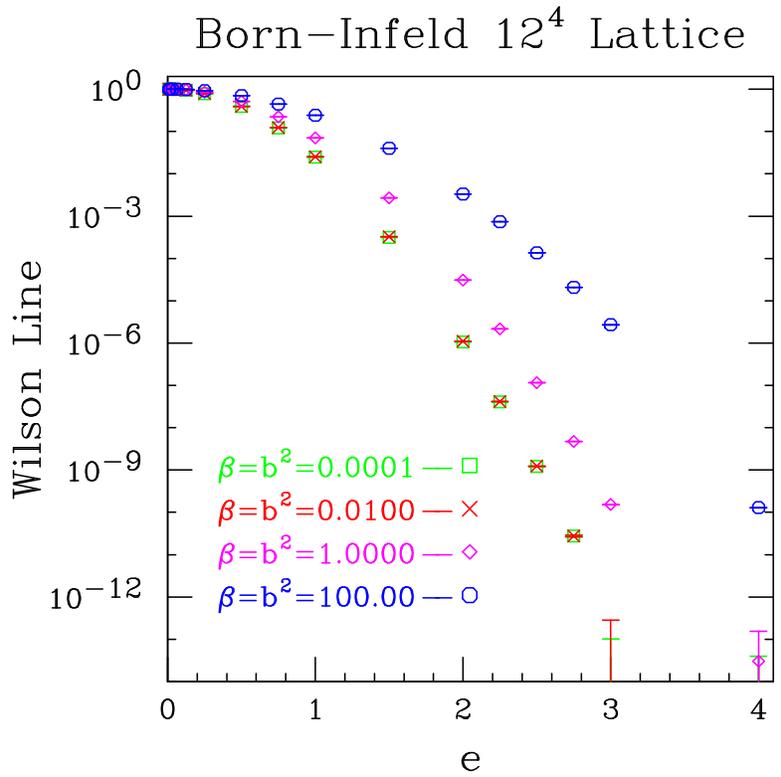}}
\caption{Wilson Lines as functions of charge $e$, for a range
of $\beta=a^4 b^2$ values: a) $8^4$ lattice: b) $12^4$ lattice.}
\label{fig:wilson}
\end{figure}

Figure~\ref{fig:wilson} shows the Wilson Lines as functions of electric charge
$e$ for 16 values of this charge ranging from $0.0078125$ to $4.0$ and for
the 4 beta values $\beta=0.0001,0.01,1.0,100.0$. At each $\beta$, 
$\langle W \rangle$ starts from $1$ at $e=0$, and falls faster than 
exponentially with charge $e$ as $e$ is increased. The falloff becomes faster
as $\beta$ (or $b$) is decreased, and the field theory becomes more non-linear.
We note, however, that the values of the Wilson lines for $\beta=0.0001$ and
$\beta=0.01$ are almost identical. This is further evidence that the theory
is close to the $\beta=0$ theory for $\beta \lesssim 0.01$. This also gives us
the first evidence for differences between the electrostatics of the quantum
and classical theories. Classical electrostatics has ${\bf B}$ and ${\bf H}$
identically zero, the contribution of equation~\ref{eqn:limit} vanishes and
the argument the square root of equation~\ref{eqn:Sminkowski} reduces to
$1-b^{-2}\bm{E}^2$ for all $b$. In the quantum theory it is not even
consistent to set ${\bf B}=0$, and the result is that the action still reduces 
to the $b$($\beta$) independent form of equation~\ref{eqn:limit} as 
$b \rightarrow 0$. We note that the L\"{u}scher-Weisz prescription yields an
excellent signal for the Wilson line with small error bars over almost 11
orders of magnitude.

We measure the ${\bf E}$ field for on axis separations (for definiteness 
consider this to be in the $z$ direction) from the electric charge. Here we
need only measure $E_z$. For this we need to measure plaquettes in the $(z,t)$
plane containing the Wilson Line. We call the $z$ separation of the centre of
such a plaquette from the Wilson line $Z$. Clearly $Z$ takes the values
$\frac{1}{2},\frac{3}{2},\frac{5}{2},...,N_z-\frac{1}{2}$. It is also clear
that
\begin{eqnarray}
\langle E_z(Z)\rangle &=& -\langle E_z(N_z-Z)\rangle    \nonumber    \\
\langle D_z(Z)\rangle &=& -\langle D_z(N_z-Z)\rangle.
\end{eqnarray}
The lattice $D_z$ field is obtained by calculating 
$\frac{\partial S}{\partial E_z}$, where $E_z$ is a $(z,t)$ plaquette and $S$
is the lattice action. The $\langle E_z \rangle_\rho$ and $\langle D_z
\rangle_\rho$ are obtained using equation~\ref{eqn:expect}. Again, the
L\"{u}scher-Weisz method yields a good signal. In what follows, we will drop
the subscripts $\rho$ and $z$ for convenience.

We will start by considering ${\bf D}$. Since the equation of motion 
$\bm{\nabla \cdot D}=\rho$ holds for all $\beta$, this means that, to the
extent that we have rotational invariance, $\langle{\bf D}\rangle$ should be
independent of $\beta$, and equal to the classical value. Of course, since
we are on the lattice, rotational invariance is at best approximate, and we
expect departures from this expectation.

\begin{figure}[htb]
\epsfxsize=4in
\centerline{\epsffile{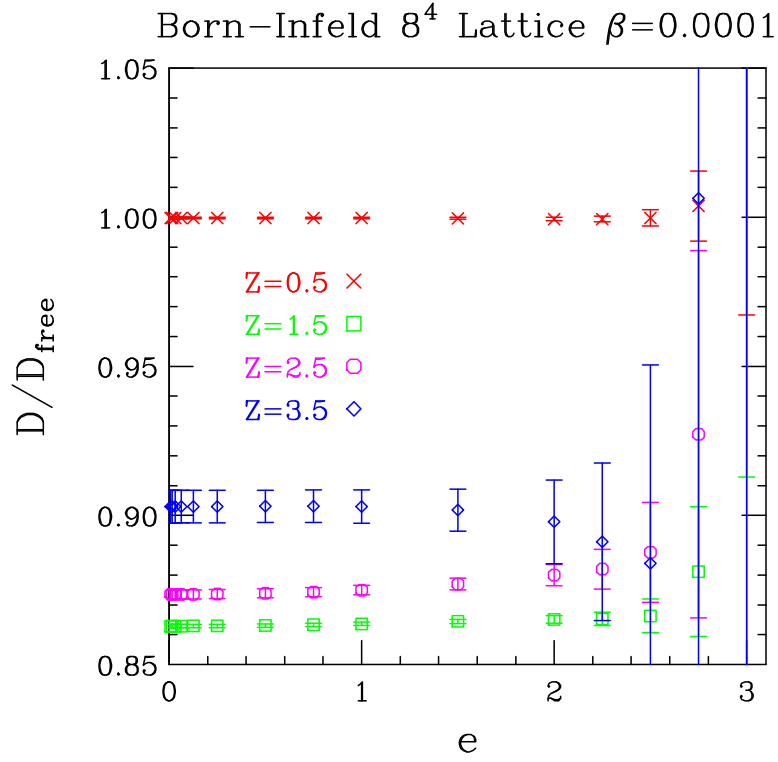}}
\vspace{0.25in}
\centerline{\epsffile{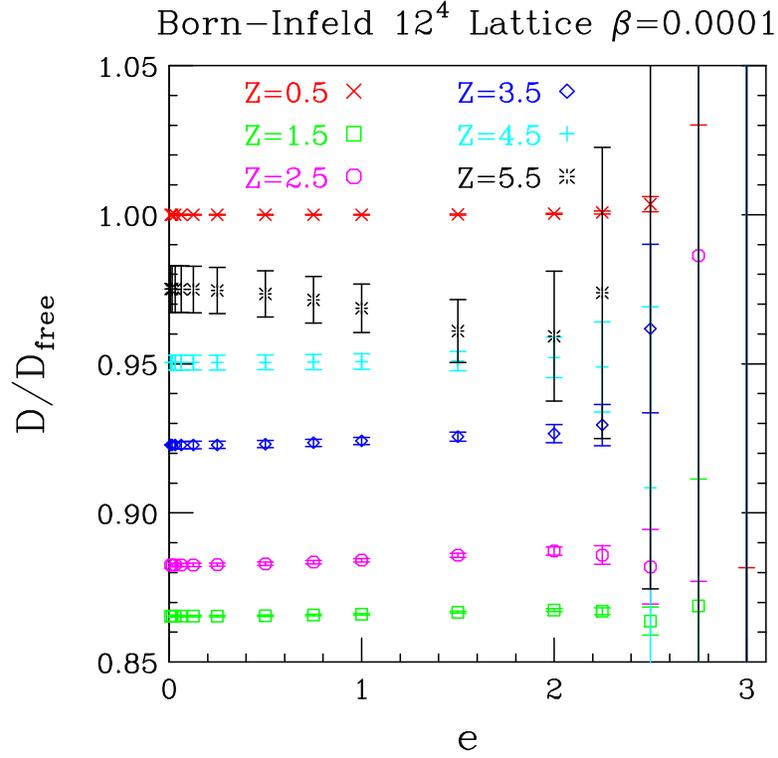}}
\caption{Ratios of ${\bf D}$ to their Maxwell (free field) values for 
$\beta=0.0001$ as functions of electric charge $e$, for accessible values of
the on-axis separation $Z$: a) for an $8^4$ lattice, b) for a $12^4$ lattice.}
\label{fig:d0.0001}
\end{figure}

Figure~\ref{fig:d0.0001} shows the ratios of the ${\bf D}$ fields produced by
a point charge $e$ on both $8^4$ and $12^4$ lattices for $\beta=0.0001$, where
we expect the action to closely approximate the small $\beta$ limit of
equation~\ref{eqn:limit}, to their free field values. For the minimum
separation of source and field ($Z=\frac{1}{2}$) $D$ is consistent with its
free-field (Maxwell) value, evaluated on the same-size lattice. This is
because, at this separation, the discrete Gauss' equation plus cubic symmetry
is sufficient to uniquely determine $D$; in fact $D=\frac{1}{6}(1-1/{\cal
V})e$. At separation $Z=\frac{3}{2}$, $D$ has fallen to $86$--$87\%$ of its
free-field value. Gauss' law and cubic symmetry are no longer sufficient to
determine the $D$ field uniquely, and we do not have the simple relation
$\bm{D} = \bm{E}$ and hence $\nabla\times\bm{D}+\partial\bm{B}/\partial t=0$
of the Maxwell theory to help us. It is thus reassuring that ${\bf D}$ is as
close to its free-field value as it is. As the separation increases, the ratio
of $D$ to its free-field value appears to be approaching $1$. This is an
indication that rotational symmetry is restored at length scales much larger
than the lattice spacing. The fact that the ${\bf D}$ for this highly
non-linear theory appears identical to that of the Maxwell theory, apart from
lattice artifacts, is non-trivial. For the Maxwell theory, $\bm{D} = \bm{E}$,
while in the $\beta\rightarrow 0$ limit 
$\bm{D}=\varepsilon(\bm{E\cdot B})\bm{B}$, where $\varepsilon$ is the familiar 
sign function.

\begin{figure}[htb]
\epsfxsize=4in
\centerline{\epsffile{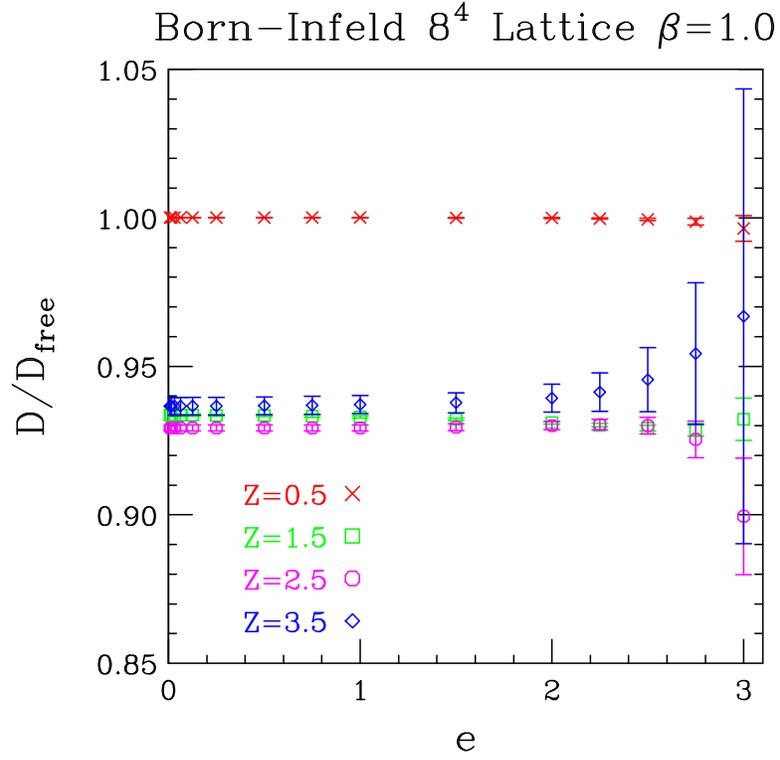}}
\vspace{0.25in}
\centerline{\epsffile{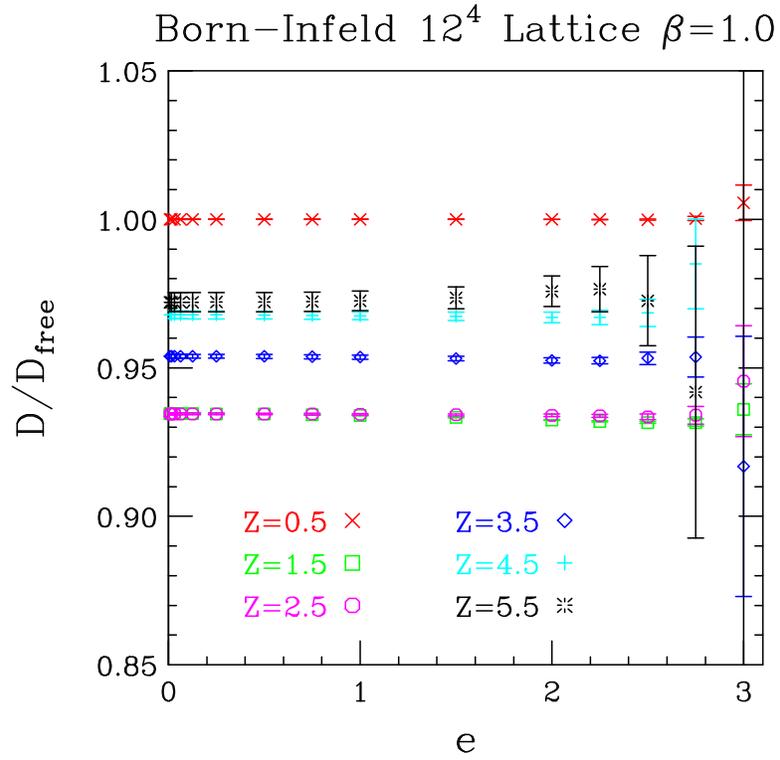}}
\caption{Ratios of ${\bf D}$ to their Maxwell (free field) values for
$\beta=1.0$ as functions of electric charge $e$, for accessible values of
the on-axis separation $Z$: a) for an $8^4$ lattice, b) for a $12^4$ lattice.}
\label{fig:d1.0}
\end{figure}

Figure~\ref{fig:d1.0} shows the $e$ dependence of $D$, for $Z < N_z/2$, for
$\beta=1.0$. This $\beta$ is close to that which maximizes 
$\langle S \rangle/V$. Thus, for this value of $\beta$, Born-Infeld QED should
be far from regular (Maxwell) QED, and also far from the non-linear limit of
equation~\ref{eqn:limit}. Again we see that at minimum separation, $D$ is
at its free field value. The ratio of $D$ to its free-field value falls to
a minimum for $1.5 \lesssim Z \lesssim 2.5$, after which it begins to increase
again. For the $8^4$ lattice, the finite lattice extent tends to suppress this
increase while the larger $12^4$ lattice shows a clear increase. Hence we see
evidence that rotational symmetry is being restored at large distances, just as
for small $\beta$. Again we believe that the difference between the $\beta=1.0$
${\bf D}$ fields and their free-field counterparts is a lattice artifact.

\begin{figure}[htb]
\epsfxsize=4in
\centerline{\epsffile{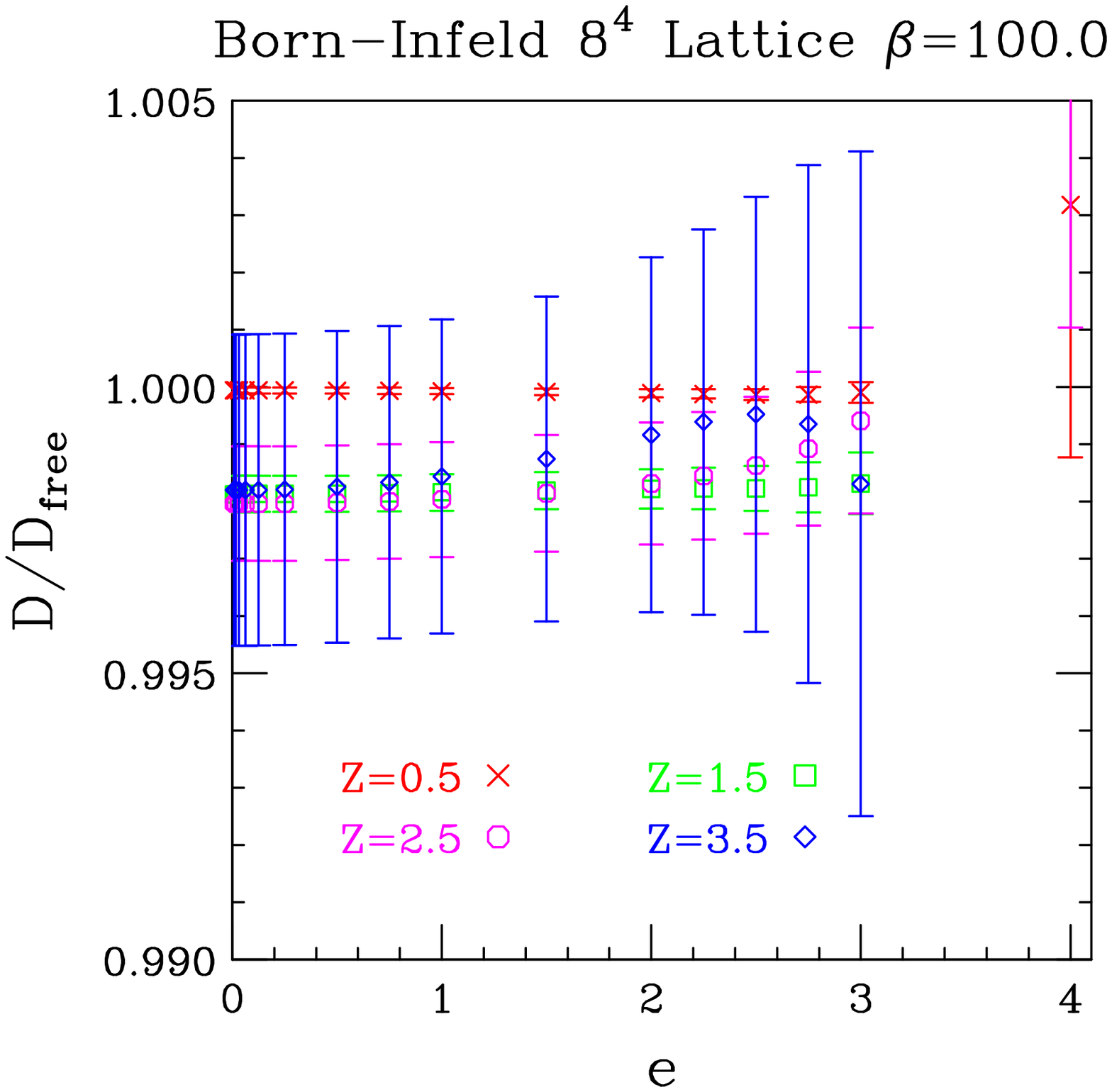}}
\vspace{0.25in}
\centerline{\epsffile{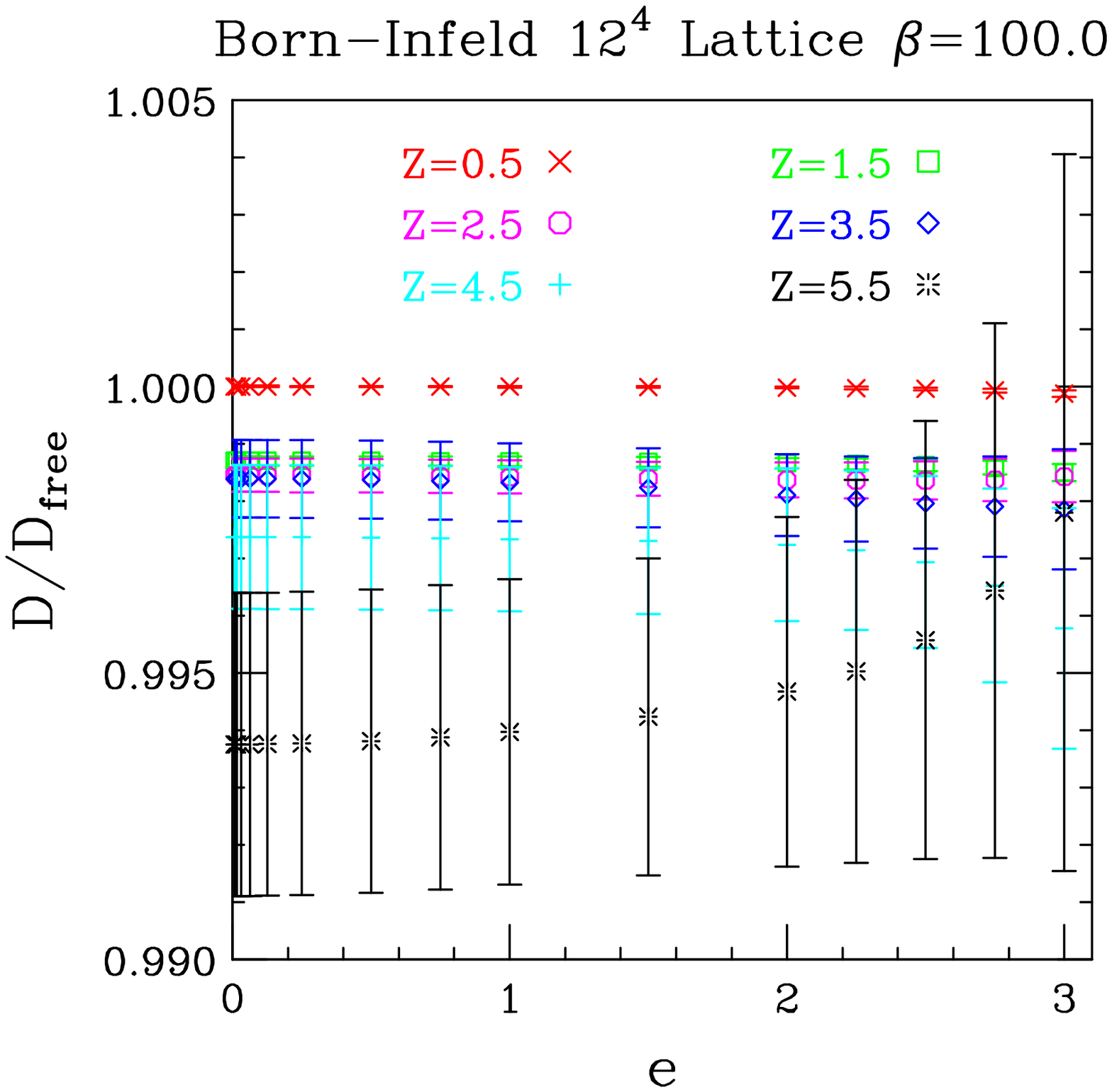}}
\caption{Ratios of ${\bf D}$ to their Maxwell (free field) values for    
$\beta=100.0$ as functions of electric charge $e$, for accessible values of     
the on-axis separation $Z$: a) for an $8^4$ lattice, b) for a $12^4$ lattice.}
\label{fig:d100.0}   
\end{figure}

In figure~\ref{fig:d100.0} we graph the ratios of $D$ to its free field value
for $\beta=100.0$, where we expect Born-Infeld QED to closely approximate
standard (Maxwell) QED. We note again that at minimum separation $D$ is 
consistent with its free field value. At larger separations there are small
but significant departures from the Maxwell theory. However, since these
departures are always less than (and often considerably less than) one percent,
this is further evidence that $\beta=100.0$ Born-Infeld QED is close to 
standard QED. Because these departures comparable in size with our error-bars,
it is difficult to make any more quantitative observations, although it does
appear that the $8^4$ and $12^4$ values are consistent for the ranges of $Z$
which they share.

We now turn our attention to the ${\bf E}$ field. Here, unlike the case of the
${\bf D}$ field, Maxwell's equations are of little help, because the relations
of equations~\ref{eqn:DH} are highly non-linear. What we are interested in
knowing is how similar these ${\bf E}$ fields are to their classical
counterparts and, in particular, whether they show the short-distance
screening of classical Born-Infeld electrodynamics. We shall use our
experience with the ${\bf D}$ fields as a guide to how much the results that
we observe are affected by lattice artifacts.

\begin{figure}[htb]
\epsfxsize=4in
\centerline{\epsffile{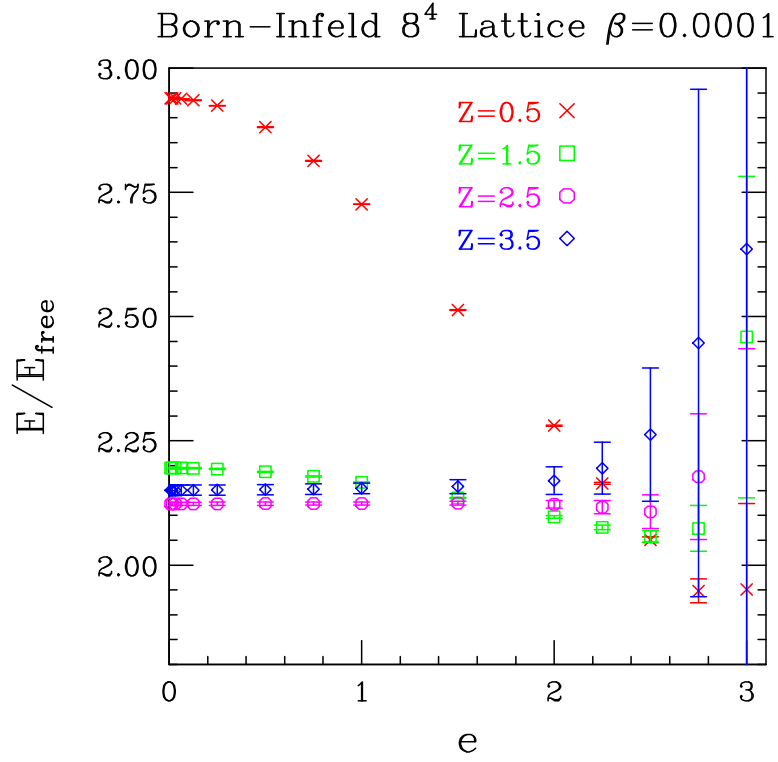}}
\vspace{0.25in}
\centerline{\epsffile{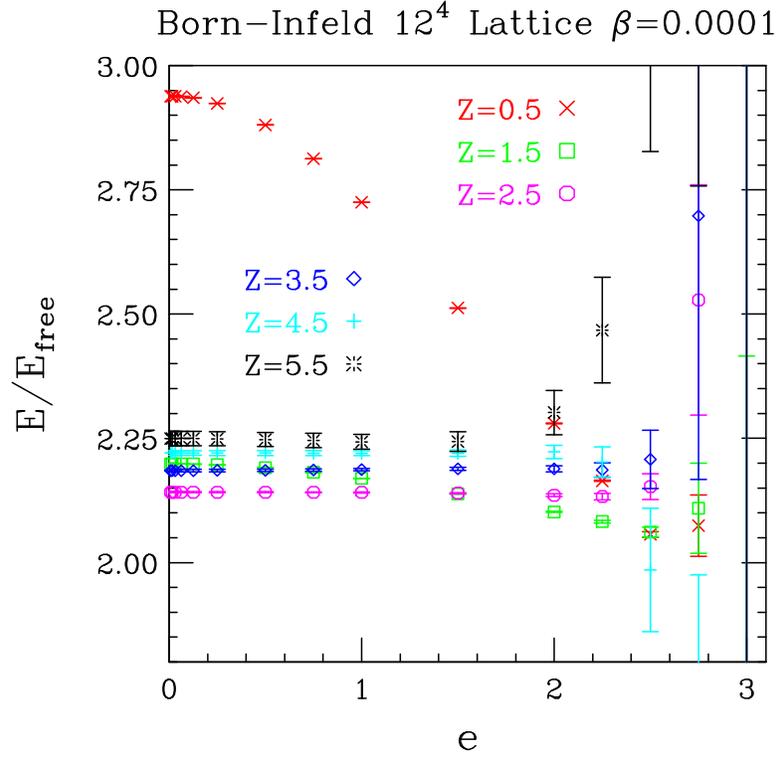}}
\caption{Ratios of ${\bf E}$ to their Maxwell (free field) values for
$\beta=0.0001$ as functions of electric charge $e$, for accessible values of
the on-axis separation $Z$: a) for an $8^4$ lattice, b) for a $12^4$ lattice.}
\label{fig:e0.0001}
\end{figure}

First we look at the case where $\beta=0.0001$ which we expect to be close
to the non-linear limit of equation~\ref{eqn:limit}. Figure~\ref{fig:e0.0001}
shows the ratios of $E$ to its free-field value for this $\beta$ as 
functions of $e$ for all accessible values of $Z$. The similar graph (not shown)
for $\beta=0.01$ is almost identical, which justifies our claim that we are
close to the non-linear limit. We first note that the $8^4$ and $12^4$ graphs
are very similar for those $Z$ values common to each, indicating that we should
be able to extract information which is not dominated by finite volume effects
from lattices of this size. Secondly we notice that for all accessible $Z$s and
over most of the range of charges considered ($0 < e \le 3$), the electric
fields are larger than their free field counterparts by a factor of 2--3. This
contrasts with the classical case where $E \le D$. This enhancement of $E$ is
due to quantum fluctuations which are clearly quite large. At $Z=\frac{1}{2}$,
it is clear that ${\bf E}$ is screened, and that this screening increases with
increasing $e$. The $Z=\frac{3}{2}$ field also shows such screening, but the
falloff with increasing $e$ is much less pronounced than at $Z=\frac{1}{2}$, 
while there is no clear signal for screening for larger $Z$. This behaviour is
qualitatively similar to what is observed classically. However, classically
its screening is determined by the ratio $e/b$ (or $r_0$), whereas in the
quantum theory, for small enough $b$ (and hence $\beta$) the screening becomes
$b$ independent. Since the limit $\beta \rightarrow 0$ can be achieved by 
taking $a \rightarrow 0$ at fixed $b$, rather than $b \rightarrow 0$, the
non-linear limit of equation~\ref{eqn:limit} could be expected to describe the
continuum limit. The question arises as to whether this limiting field theory
is trivial or not. If it is trivial, the ${\bm E}$ field of a point charge (a
propagator of this field theory) would be proportional to the free field
propagator. Of course, this would only be true of this propagator at finite
distances in physical units, which as $a \rightarrow 0$ means large lattice
distances. Thus the fact that this ratio on the $12^4$ lattice varies by only
of order $5\%$ for $Z > 0.5$, at least for small charges, could be an 
indication that this limit is trivial. 

\begin{figure}[htb]
\epsfxsize=4in
\centerline{\epsffile{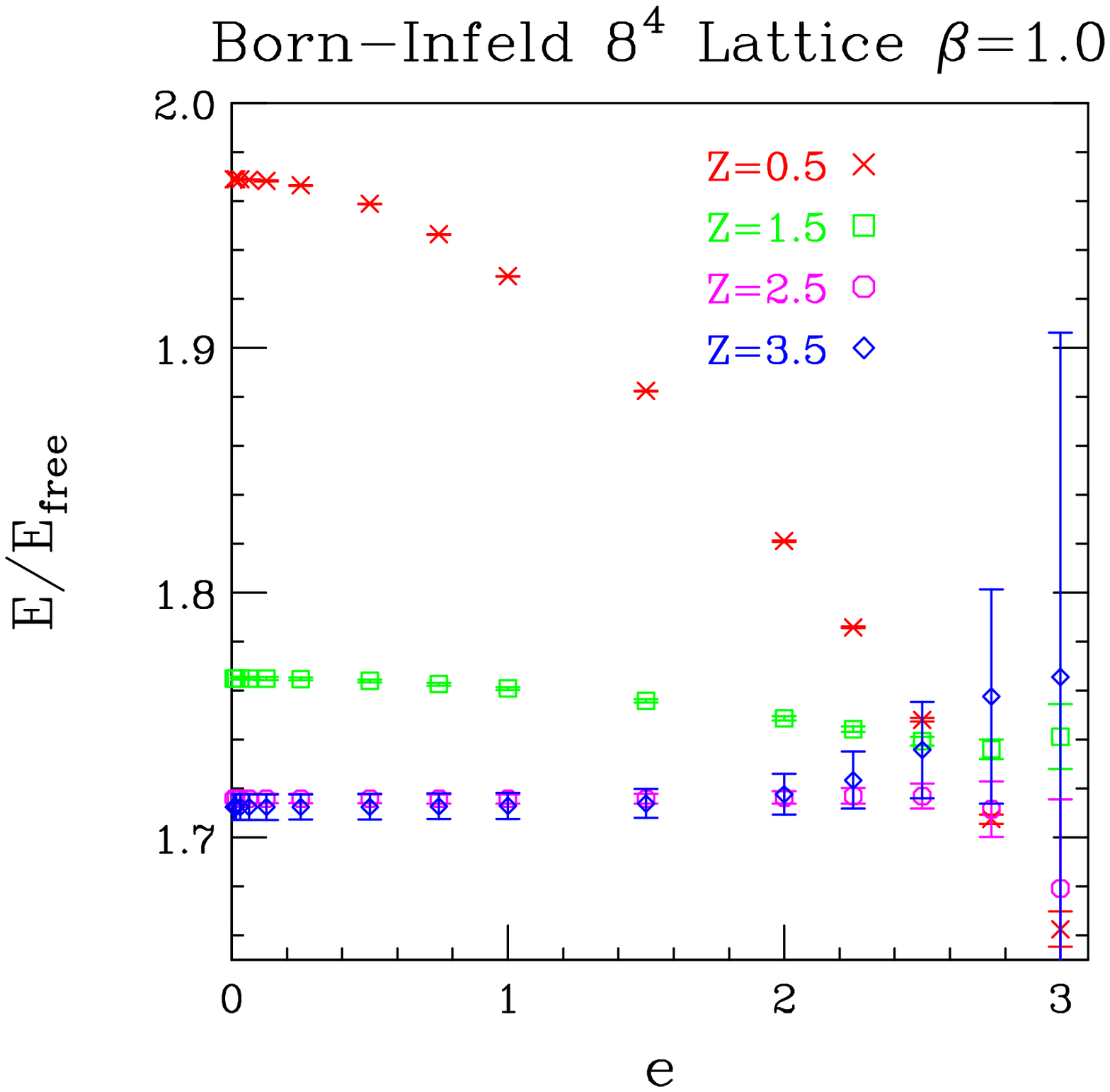}}
\vspace{0.25in}
\centerline{\epsffile{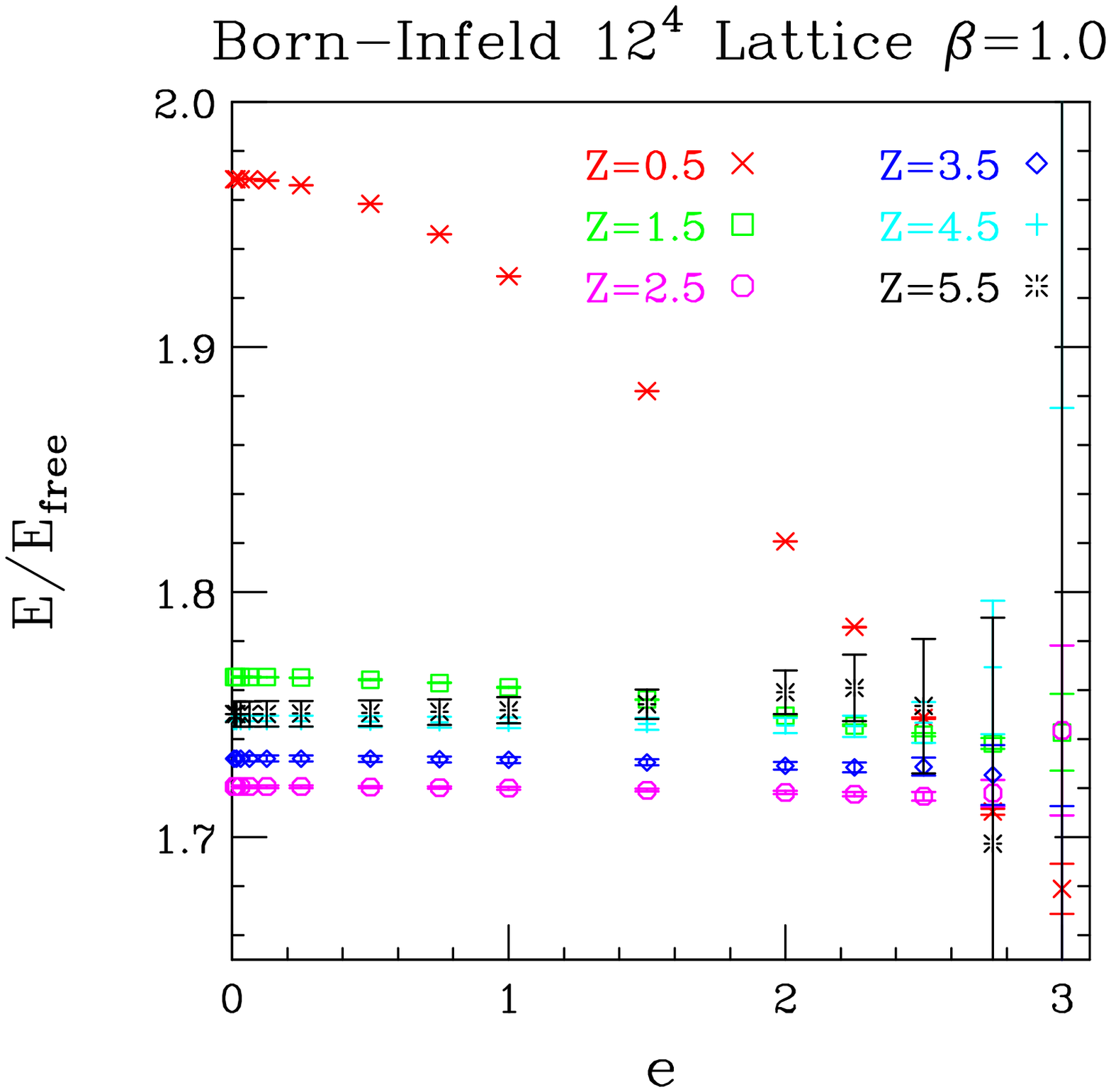}}
\caption{Ratios of ${\bf E}$ to their Maxwell (free field) values for
$\beta=1.0$ as functions of electric charge $e$, for accessible values of
the on-axis separation $Z$: a) for an $8^4$ lattice, b) for a $12^4$ lattice.}
\label{fig:e1.0}
\end{figure}
 
Figure~\ref{fig:e1.0} shows $E/E_{free}$ as functions of $e$ for all accessible 
separations $Z$, for $\beta=1.0$. Again, we note the similarity between the
results from the $8^4$ and $12^4$ lattices up until $Z$ approaches $N_z/2$
where the effect of the finite volume becomes apparent, which gives us
confidence that the finite volume effects are under control. Despite the fact
that we are now far from both the small and large $\beta$ limits, these graphs
are qualitatively similar to those for $\beta=0.0001$, except for the scale
on the vertical axis. The enhancement of this electric field over its Maxwell
value due to quantum fluctuations is smaller than that for $\beta=0.0001$.
The screening which we observe at $Z=\frac{1}{2}$ (and $Z=\frac{3}{2}$) is also
less by roughly a factor of 3 over the same range of charges $e$.

\begin{figure}[htb]
\epsfxsize=4in
\centerline{\epsffile{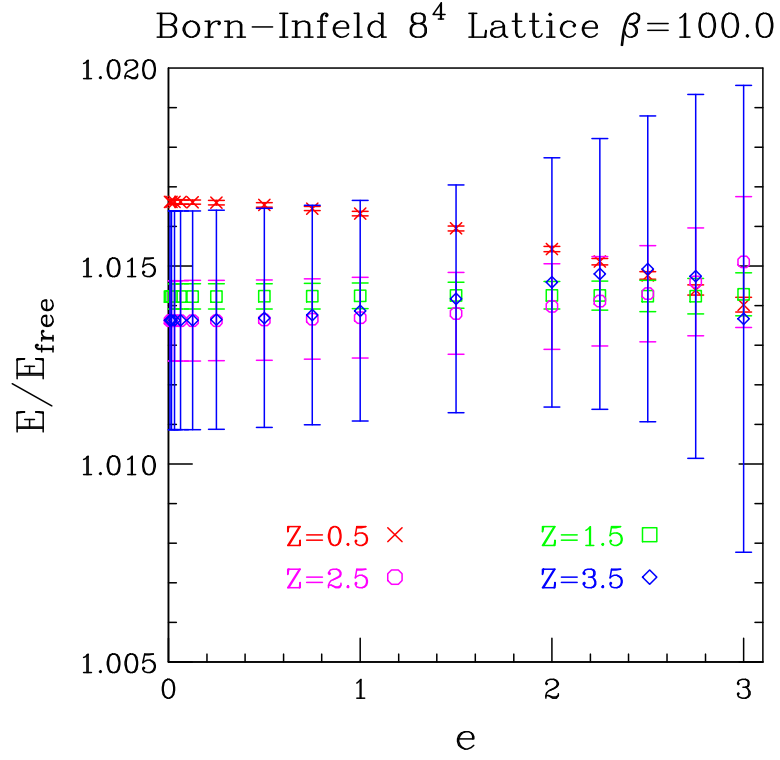}}
\vspace{0.25in}
\centerline{\epsffile{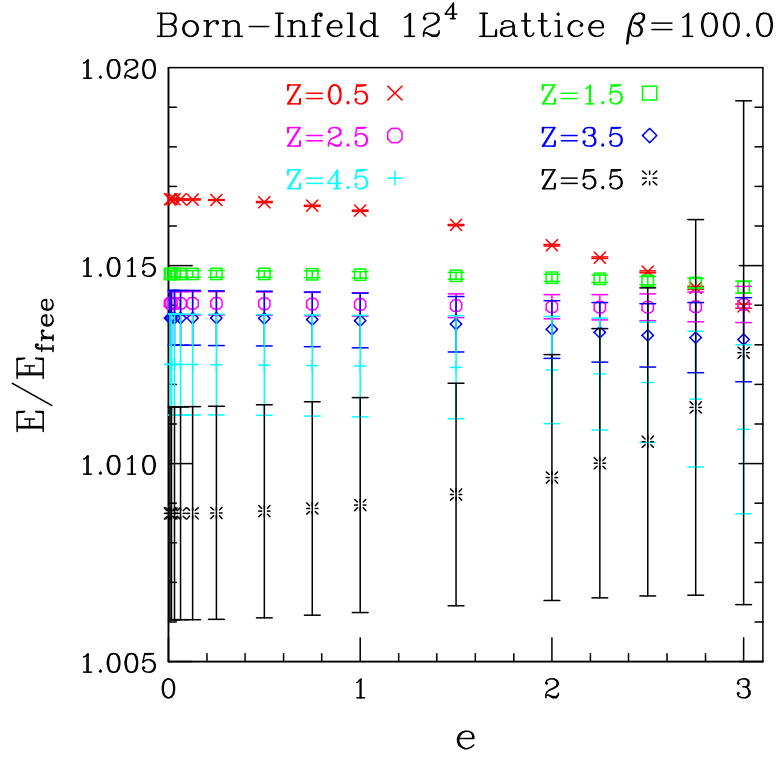}}
\caption{Ratios of ${\bf E}$ to their Maxwell (free field) values for
$\beta=100.0$ as functions of electric charge $e$, for accessible values of
the on-axis separation $Z$: a) for an $8^4$ lattice, b) for a $12^4$ lattice.}
\label{fig:e100.0}
\end{figure}

In the last of this series of figures showing the $e$ and $Z$ dependence of
the ratio of $E$ to its free field value, figure~\ref{fig:e100.0}, we show
the results for $\beta=100$ where we expect Born-Infeld QED to be close to
its Maxwell (standard QED) limit. What is clear from these graphs is that,
since the $E$ never differs from its free field value by more than 2\% over the
range of $e$ considered, Born-Infeld QED is close to its free-field limit for
$\beta=100$. We do, however, observe a small increase in $E$ over its 
free-field value due to quantum fluctuations, and see clear evidence for 
weak screening for $Z=\frac{1}{2}$. This is good evidence that expanding the
square root in the action in powers of $1/\beta$, and using a perturbative
analysis, would be valid in this domain.

\begin{figure}[htb]
\epsfxsize=4.1in
\centerline{\epsffile{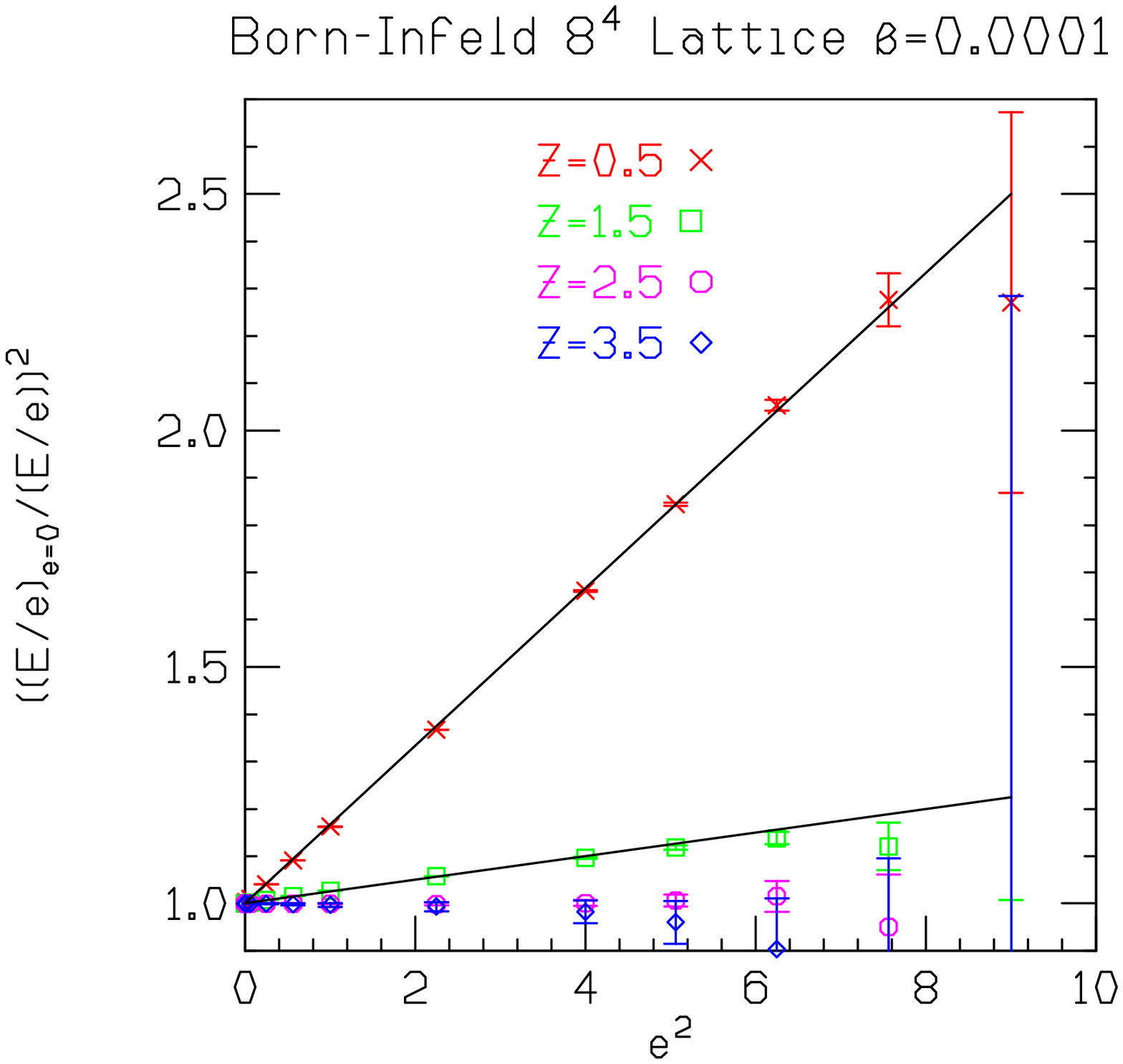}}
\vspace{0.25in}
\centerline{\epsffile{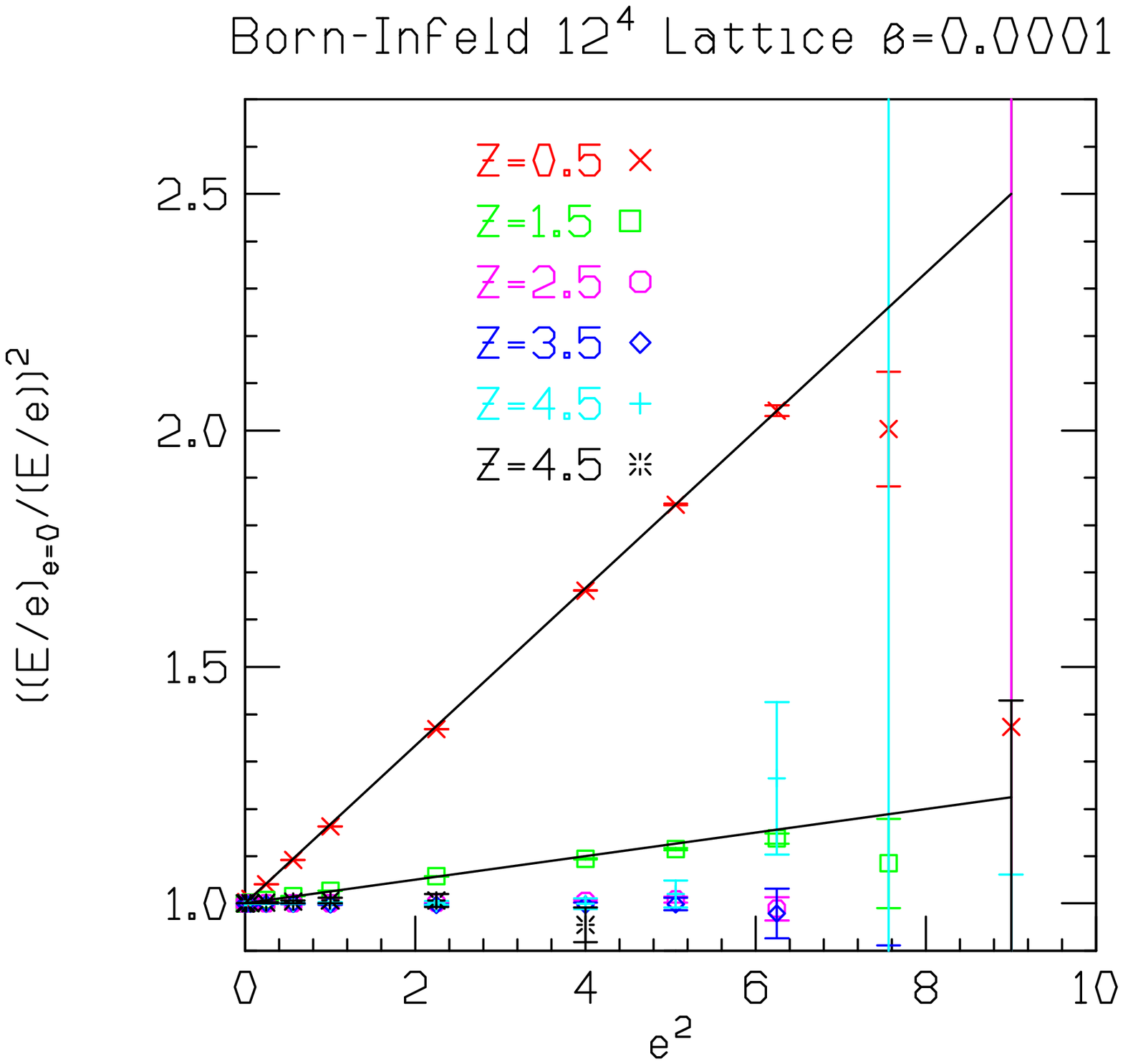}}
\caption{$e^2$ dependence of the inverse electric field: a) on an $8^4$ lattice
b) on a $12^4$ lattice. Note that the straight lines drawn to aid the eye on
(a) and (b) are identical.}
\label{fig:screening}
\end{figure}

In an effort to get a more quantitative understanding of the screening of the
${\bf E}$ field that we have observed, we look again at the screening in 
classical Born-Infeld electrodynamics. Here we can expose the screening length
($r_0$) by looking at
\begin{equation}
\left\{ {[\bm{E}(\bm{r})/e]_{e=0} \over [\bm{E}(\bm{r})/e]} \right\}^2
= 1 + \left({r \over r_0}\right)^4 = 1 + {1 \over 16\pi^2 b^2 r^4}\,e^2.
\end{equation}
So, for the classical case, plotting the left hand side of this equation for
fixed $r$ against $e^2$ (or $1/r_0^4$) would give a straight line. 
Figure~\ref{fig:screening} shows this quantity from our simulations at
$\beta=0.0001$. While straight line fits to the $Z=\frac{1}{2}$ and
$Z=\frac{3}{2}$ are impossible, the straight lines we have drawn to guide the
eye indicate that the $e^2$ dependence is almost linear. Perhaps the observed
departures are due to lattice artifacts. What this behaviour does suggest is
that the screening increases without bound as $e^2$ is increased, as is true
classically.

\begin{figure}[htb]
\epsfxsize=4.1in
\centerline{\epsffile{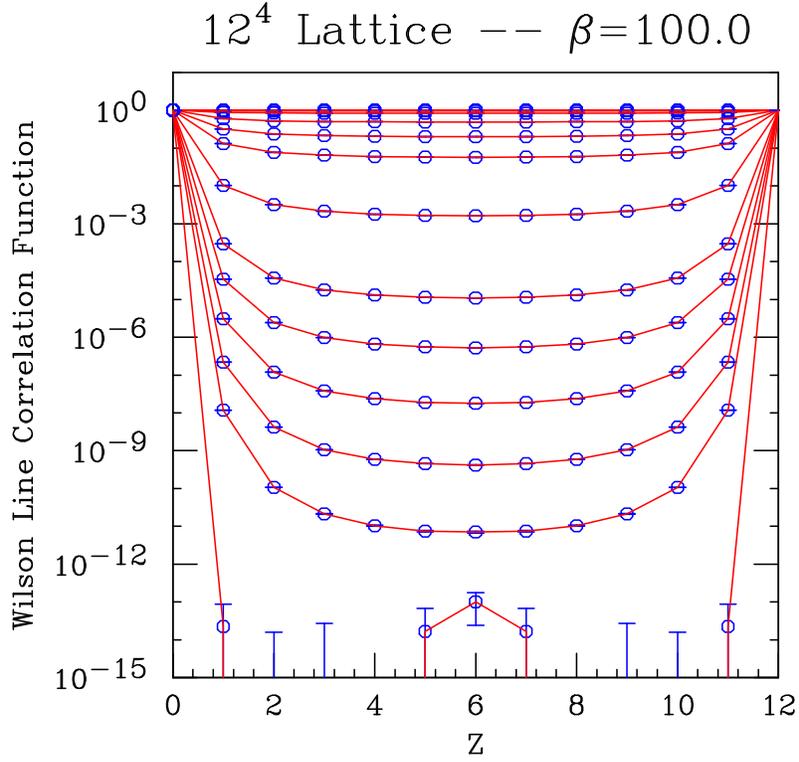}}
\vspace{0.25in}
\centerline{\epsffile{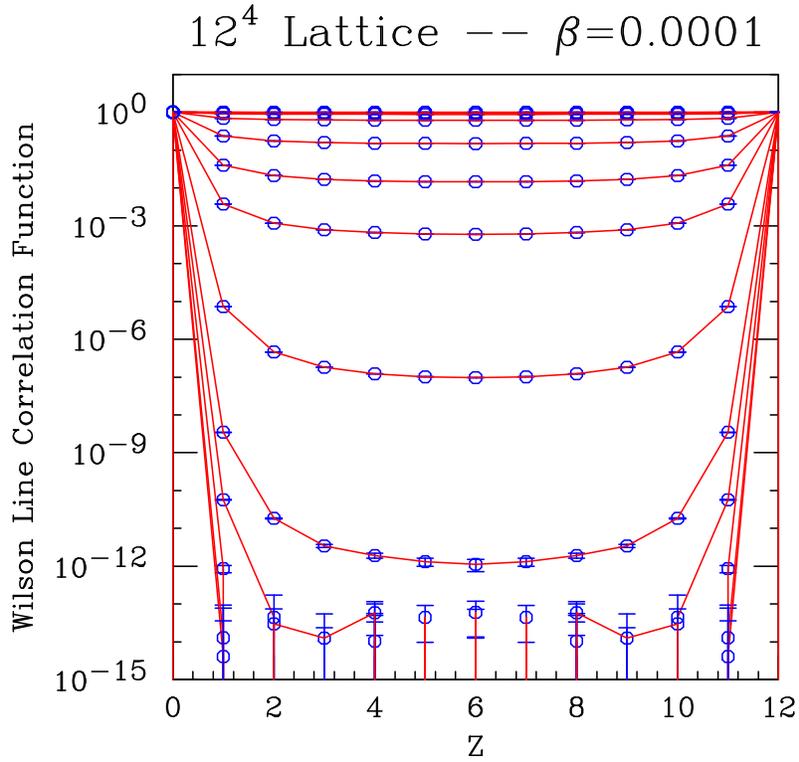}}
\caption{Wilson line correlation functions for charges $e$ ranging from
$0.0078125$ (top curve) to $4$ (bottom curve) a) for $\beta=100.0$ and
b) for $\beta=0.0001$.}
\label{fig:wilcor}
\end{figure}

Finally, for completeness, we show the Wilson Line correlations for each of
the 16 charges for $\beta=100$ and $\beta=0.0001$ 
on the $12^4$ lattice in figure~\ref{fig:wilcor}.
These correlation functions show the interactions 
between charge $+e$ and charge $-e$ separated by distance $Z$. What these
indicate is that the effect of screening on such interactions is to reduce
the correlation functions as $\beta$ is decreased (which as observed above, 
increases screening). One expects that the Wilson Line correlation function
will behave as
\begin{equation}
\langle W(\bm{x})W^\dagger(\bm{x}+\bm{Z}) \rangle = C(Z)\exp[-V(Z)T].
\end{equation}
Although the electric field shows screening, $E$ for $\beta=0.0001$ exceeds
that for the free field and hence that for $\beta=100.0$ for the same charge,
over the range of charges considered. Hence, the potential energy $V(Z)$ at
$\beta=0.0001$ exceeds that for $\beta=100.0$. It is for this reason that this
correlation function falls more rapidly for the smaller $\beta$.

\section{Discussion and Conclusions}

Euclidean Born-Infeld electrodynamics has been quantized on a discrete 
space-time lattice. The Euclidean action is positive, allowing us to simulate
this theory by importance sampling techniques. We have simulated the 
gauge-invariant non-compact lattice implementation on $8^4$ and $12^4$ lattices
using standard Metropolis Monte-Carlo methods for several values of the
non-linearity parameter $b$ [or the dimensionless $\beta=b^2a^4$ ($a$ is the
lattice spacing)]. We measure the electric fields produced by a static point
charge $e$, as functions of that charge. On the lattice this charge is
introduced by the inclusion of a Wilson line (Polyakov loop). The phase of
this Wilson line introduces a sign problem. However, this sign problem is 
overcome using the method of L\"{u}scher and Weisz, which uses partial 
factorization of the functional integral to produce the exponential statistics
required to circumvent such sign problems. With this, we were able to obtain
excellent signals for the Wilson line, ${\bf E}$ and ${\bf D}$ fields, over
an appreciable range of $e$ values.

We performed simulations with $\beta$ values ranging from those large enough
to approximate the Maxwell (free-field) theory to the those where Born-Infeld
QED is well approximated by a $\beta$($b$)-independent conformal field theory
with Euclidean Lagrangian ${\cal L}_E = |\bm{E\cdot B}|$ or (in Minkowski
space) by the Hamiltonian ${\cal H} = |\bm{D} \times \bm{B}|$. For all $\beta$
values, the electric displacement ${\bf D}$ appears to differ from its value
in the Maxwell theory only by lattice artifacts. The electric field ${\bf E}$
shows considerable enhancement over its Maxwell value by as much as a factor of
$\approx 3$ due to quantum fluctuations. Away from the Maxwell limit, ${\bf E}$
exhibits short-distance screening with a screening length whose 4th power is
approximately linear in $e^2$, as in the classical theory. Also as in the
classical theory, screening is enhanced as $\beta$($b$) is increased. However,
unlike in the classical case where the screening at fixed $e$ increases
without limit as $b$ is increased, in the quantum theory this screening
approaches a limit described by the conformal field theory mentioned at the
beginning of this paragraph. Once this limit is achieved, the only way to
increase the screening length is to increase $|e|$.

Since our chief reason for studying Born-Infeld QED is because of its
connection to strings and branes, we only need to consider it as an effective
field theory with a momentum cutoff, or in our language, a finite lattice 
spacing. However, it is interesting to consider whether one can remove the
momentum cutoff (take $a \rightarrow 0$) in such a way as to define a 
renormalizable field theory. If we fix $b$ and take $a$ to zero, we are 
taking the $\beta \rightarrow 0$ limit, where the field theory is described
by the Euclidean Lagrangian ${\cal L}_E = |\bm{E\cdot B}|$. Hence this theory
describes the infinite-momentum-cutoff limit of Born-Infeld QED. What we do
not know, is whether this quantum field theory is non-trivial. Our simulations
hint that the propagators of this theory might show free-field scaling.
However, the evidence is far from convincing, especially if we consider the
possibility that deviations from free-field scaling could well be logarithmic. 

It would be useful to perform simulations of Born-Infeld QED on larger lattices,
possibly restricting these simulations to those using the action based on
the Lagrangian ${\cal L}_E = |\bm{E\cdot B}|$. Of more interest would be
simulations using those actions obtained from higher dimensional Born-Infeld
theories by dimensional reduction, which describe strings or branes with
transverse degrees of freedom. In these actions, the square root contains an
additional term $(\partial_\mu X_a)^2$, where the scalar fields $X_a$ are the
$n-p$ additional components of the $A$ field corresponding to the reduced
dimensions and are interpreted as the transverse displacements of the
$p$-brane. We plan such simulations in the near future.

\section*{Acknowledgements}

One of us (DKS) would like to thank Cosmas Zachos for several stimulating
discussions. The $12^4$ simulations were performed on the Rachel supercomputer
at PSC under an NRAC computing allocation from the NSF. The $8^4$ simulations
were run on a Linux PC belonging to the HEP Division at Argonne National 
Laboratory.

\end{document}